# MICROTUBULE DYNAMIC INSTABILITY:
# THE ROLE OF CRACKS BETWEEN PROTOFILAMENTS

Chunlei Li,[a] Jun Li,[a] Holly V. Goodson*[b,c,d] and Mark S Alber,*[a,b,e]

[a] *Department of Applied & Computational Mathematics and Statistics, University of Notre Dame, IN, USA.*

[b] *The Interdisciplinary Center for the Study of Biocomplexity, University of Notre Dame, IN, USA.*

[c] *Department of Chemistry and Biochemistry, University of Notre Dame, IN, USA.*

[d] *Department of Biological Sciences, University of Notre Dame, IN, USA.*

[e] *Department of Medicine, Indiana University School of Medicine, IN, USA.*

*\* Authors for correspondence,* Mark S. Alber, E-mail: malber@nd.edu, Holly V. Goodson, E-mail: hgoodson@nd.edu

# ABSTRACT

Microtubules (MTs) are cytoplasmic protein polymers that are essential for fundamental cellular processes including the maintenance of cell shape, organelle transport and formation of the mitotic spindle. Microtubule dynamic instability is critical for these processes, but it remains poorly understood, in part because the relationship between the structure of the MT tip and the growth/depolymerization transitions is enigmatic. What are the functionally significant aspects of a tip structure that is capable of promoting MT growth, and how do changes in these characteristics cause the transition to depolymerization (catastrophe)? In previous work, we used computational models of dynamic instability to provide evidence that cracks (laterally unbonded regions) between protofilaments play a key role in the regulation of dynamic instability, with deeper cracks being more likely to lead to catastrophe, and disappearance of cracks promoting rescue. Here we use computational models to investigate the connection between cracks and dynamic instability in more detail. Our work indicates that while cracks contribute to dynamic instability in a fundamental way, it is not the depth of the cracks *per se* that governs MT dynamic instability. Instead it is whether the cracks terminate in GTP-rich or GDP-rich areas of the MT that governs whether a particular MT tip structure is likely to grow, shrink, or transition: the identity of the crack-terminating subunit pairs has a profound influence on the moment-by-moment behavior of the MT. Based on these observations, we suggest that a functional "GTP cap" (i.e., one capable of promoting MT growth) is one where the cracks terminate in pairs of GTP-bound subunits. In addition to helping to clarify the mechanism of dynamic instability, this idea has relevance for the mechanism of MT stabilizers: proteins that introduce lateral cross-links between protofilaments would produce islands of GDP tubulin that mimic GTP-rich regions in having strong lateral bonds, reducing crack propagation, suppressing catastrophe and promoting rescue.

1. **Introduction**

Microtubules (MTs) are components of the cytoskeleton, the set of dynamic proteinaceous polymers that give cells shape, the ability to move, and the ability to create and maintain internal organization. MTs themselves perform essential roles such as acting as train tracks for the molecular motors that catalyze the movement of various cargos and pulling the chromosomes apart during cell division. Essential to all of these functions is the fact that MTs are dynamic: they grow and shrink in a stochastic process called dynamic instability [1-4]. Dynamic instability (DI) renders to MTs the ability to explore space within the cell (allowing them to come into contact with poorly diffusible cargo like membranous organelles) and respond quickly to changes in the intracellular and extracellular environments. As is typical for biological processes, dynamic instability is regulated by a multitude of proteins that bind to MTs and alter various aspects of DI.

While the existence and importance of dynamic instability are generally accepted, the mechanism of dynamic instability is less clear. Microtubules (MTs) consist of tubulin subunits (dimers of alpha and beta tubulin) assembled into 13 linear polymers called protofilaments, which are themselves assembled into a hollow tube [1-4] (Fig. 1). Each tubulin subunit binds the energy carrier GTP (similar to the more familiar ATP), which is hydrolyzed (chemically broken into GDP and free phosphate) a short time (on the order of a second or less) after polymerization. The textbook explanation of dynamic instability posits that GTP-bound tubulin subunits are in a conformation (shape) that is relatively straight and is capable of forming strong lateral bonds with other GTP tubulin subunits, leading to stable MT structures. In contrast, GDP tubulin is preferentially bent and is incapable of forming or maintaining strong lateral bonds. Thus, as long as a MT maintains a "cap" composed of GTP-bound subunits (a GTP cap), it can grow, but when it loses its GTP cap through GTP hydrolysis, the action of regulatory proteins, or other mechanisms, the MT falls apart [1, 3-6].

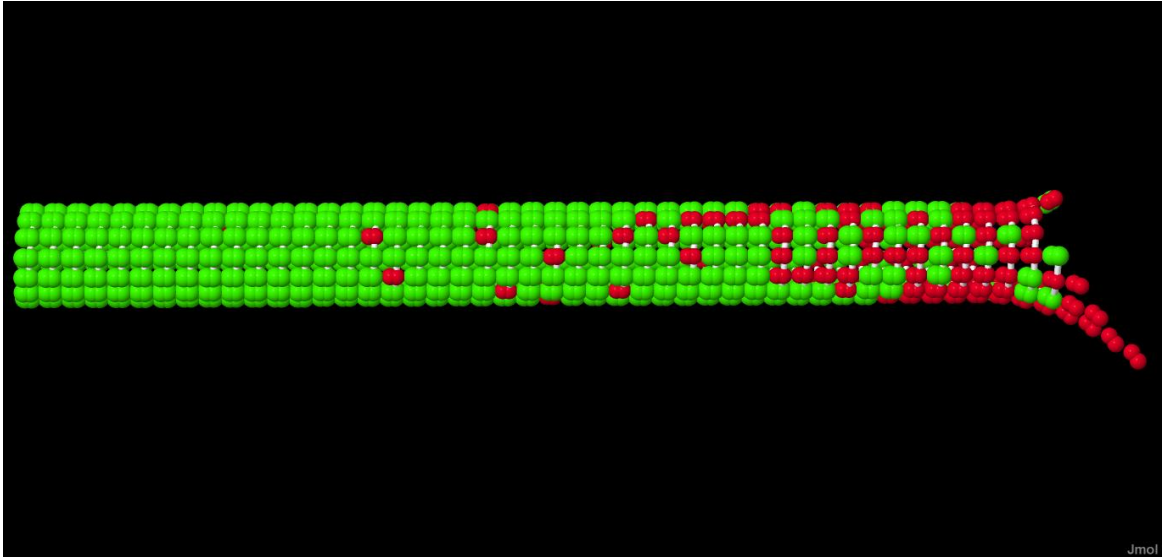

**Fig. 1.** Three dimensional representation of a typical microtubule tip structure that occurred in the simulated microtubules during the growth phase (10 μM tubulin). The left end is the minus end and the right end is the plus end. Green and red subunits represent GDP and GTP-bound tubulin subunits respectively, while the white bars represent lateral bonds.

While this conceptual model is widely accepted, many aspects of it remain undefined. For example, how big is the GTP cap? What is its shape? What causes a MT that is in one state (growth or depolymerization) to transition to the other? More precisely, what are the functionally significant aspects of a cap structure that is capable of promoting MT growth, and how do changes in these characteristics lead to catastrophe? Because it is very difficult to answer these questions experimentally, we turned to computational modelling. Our goal was to build a computational model that incorporates the essential elements of established MT structure (such as lateral bonds) and biochemistry, while being fast enough to simulate single MTs or even systems of competing MTs over experimentally relevant spans of time. In previous work, we described such a model, and used it to investigate the structure of the MT cap and the mechanism of dynamic instability [11, 12]. We provided evidence that the top region of MTs is not completely laterally bonded, and defined the regions where neighboring protofilaments lacked lateral bonds as "cracks." Also, stathmin was recently incorporated into this model and the various activities of stathmin were tested in [14].

Using a mean-field approximation of the computational model [12], we found that the depth of the inter-protofilament cracks correlates with the behavior of the MT, suggesting that crack depth plays a fundamental role in dynamic instability. However, the results with the computational model itself were less clear: while the crack depth does fluctuate rapidly in this model, and the average crack depth differs between the growth and depolymerization states, analysis of crack depth was not able to predict an incipient transition [11]. Instead, we observed that catastrophe tended to occur when one or more cracks extended into GDP-rich regions, and that rescue became more likely when pairs of the GTP subunits that rapidly exchange on depolymerizing MT tips are able to establish lateral bonds [11].

In this paper, we use the computational model to provide a more thorough investigation of the connection between the structure of the MT tip and dynamic instability activity. Consistent with our previous work with the mean field model [12], we show that altering the depth of the inter-protofilament cracks in otherwise identical MT tips does influence MT behavior. However, contrary to our expectations, we find that while extending the cracks in growing MTs does increase the likelihood of catastrophe, healing the cracks (i.e., zipping up all lateral bonds between protofilaments) in depolymerizing MTs has relatively little effect. Instead, we show that altering the identity of the subunit pairs at which the cracks terminate has a dramatic influence on MT behavior. Moreover, multinomial logistic regression models that are based on evaluation of the computational model and that use as predictors the identity of the crack termination subunits and the size of the GTP cap are able to predict the likelihood that particular spontaneously occurring tip structures will lead to growth or depolymerization. These observations demonstrate that it is not the depth of the cracks *per se* that matters in determining whether a MT is likely to grow, shrink, or undergo transition, but whether the cracks terminate in the regions of GTP or GDP-rich tubulin.

On the basis of these observations, we suggest that a growth promoting "GTP cap" is one where the cracks terminate in pairs of GTP-bound subunits. Note that this definition is relevant only to a particular moment in time, since the MT tip fluctuates rapidly. However, it follows that MTs will tend to undergo period of extended growth under conditions where the cracks are likely to terminate in pairs of GTP-bound subunits, and unlikely to terminate in GDP subunits. Thus, for dynamic MTs, the depth (size) of the GTP cap is also important because when the cap is deep, fluctuations in the depth of the cracks are less likely to lead to crack extension into GDP-rich regions. These ideas lead us to propose that a functionally significant (growth promoting) GTP cap is, on average, one where it is unlikely that cracks extend into GDP-rich regions. In addition to helping to clarify the mechanism of dynamic instability, this idea has relevance for the mechanism of MT stabilizers: proteins that introduce lateral cross-links between protofilaments would produce islands of GDP tubulin that mimic GTP-rich regions in having strong lateral bonds, reducing crack propagation, suppressing catastrophe and promoting rescue.

## 2. Methods

### 2.1. Description of the computational model

All simulations in this study use the computational model developed in Margolin et al. [11, 12]. Briefly, the MT in this model consists of 13 protofilaments with explicit lateral bonds between neighboring protofilaments. The model involves five possible events: the attachment of a GTP-bound subunit on the top of an individual protofilament; the detachment of a subunit (or subunits) from the top of individual protofilament (which is only possible for subunits that are not laterally bonded); the formation of a lateral bond between two subunits from two neighboring protofilaments; the breakage of a lateral bond between two neighboring subunits (which is only possible if this subunit pair is the topmost lateral bond); and the hydrolysis of a GTP-bound subunit to GDP-bound subunit. As noted above, by "subunit" here we mean individual tubulin dimer. The waiting times of these events are represented by exponential random variables. For a more detailed description of this model, please refer to [11, 12].

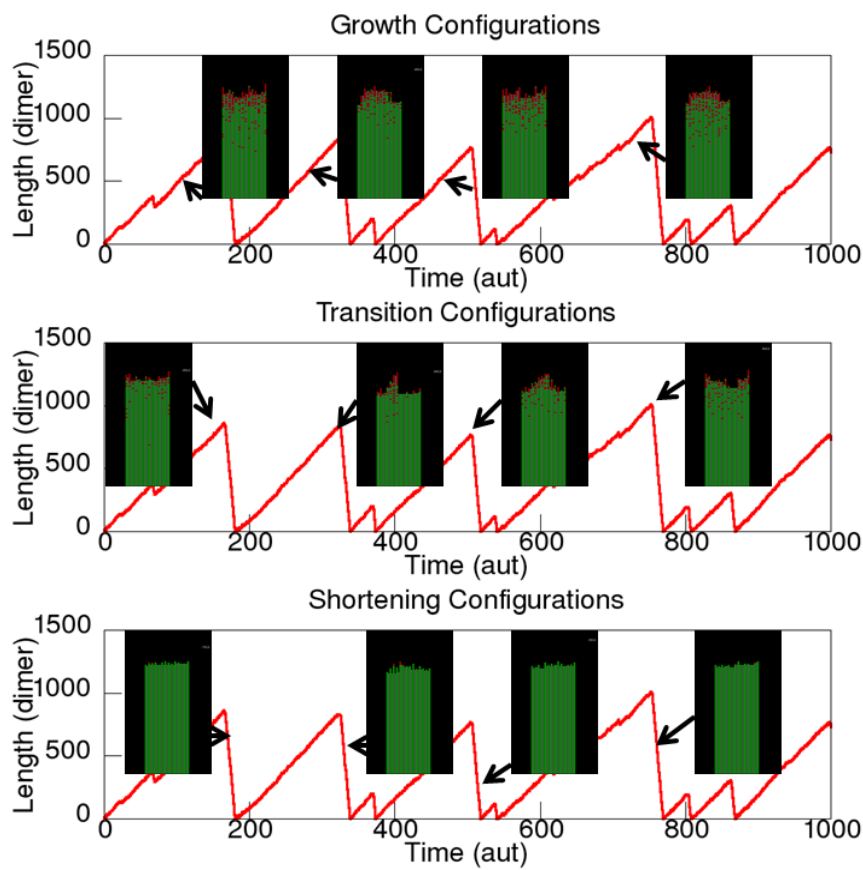

**Fig. 2.** Reference structures extracted from the initial simulation and used for further analysis (see Methods and [11]). The x-axis is time in arbitrary units of time (aut), which can be approximated as seconds. The y-axis is the length of MT. The top panel shows the four growth configurations, the middle the transitional configurations, and the bottom the shortening configurations. Green boxes signify GDP-bound subunits while red boxes signify GTP-bound subunits.

### 2.2. Data acquisition

*Reference simulation and reference structures.* An initial simulation was performed using our standard parameter set (set C from reference [11]) at 10 μM free tubulin (constant concentration, length-history plot for this reference simulation is shown in Fig. 2). Twelve specific configurations (four each for growing, shrinking, and transitioning cases) were arbitrarily chosen from this simulation for further study and are referred to below as "reference structures." Images of these reference structures and their relationship to the behavior of the MT are provided in Fig. 2.

*Procedure for testing effect of altering the crack depth.* To test the effect of altering the crack depth, we changed the crack depth as described in the relevant figure (e.g., by healing all cracks or extending the cracks three subunits deeper) for each of the reference structures shown in Fig. 2, and then ran a series of 20 new simulations using the altered structure as the initial configuration. In parallel, we ran 20 simulations using the original structure as the initial configuration. We then repeated this procedure across a range of different tubulin concentrations and used obtained data to calculate the probability of growth/shortening as a function of the tubulin concentration for each reference structure and its altered forms.

To calculate the probability from the simulation data of each configuration growing or shortening, we defined the MT to be growing if its average length increased more than 25 subunits in the first 5s and depolymerizing if its average length decreased more than 150 subunits in the first 5s., as in our previous work [11]. The MT was assumed to be in transition if it failed to meet either the growing or shortening conditions. Notice that in each case when a reference structure was altered by healing or extending the cracks, it was altered only in the initial configuration. All simulations starting from these altered initial configurations were performed using otherwise standard rules.

*Procedure for testing the effect of altering the subunits at the bottom of the cracks.* This approach is similar to that used to assess the effect of altering crack depth. Briefly, we identified the subunit pair at the bottom of each crack, changed each subunit to either GTP or GDP tubulin subunit as appropriate, and then ran a set of 20 new simulations using the altered structure as an initial configuration. This procedure was repeated for a range of tubulin concentrations. The resulting simulation data were used for calculating the probability of growth/shortening as a function of tubulin concentration for each reference structure and its altered form. All simulations starting from these altered configurations were run using the standard rules unless otherwise indicated (Fig. 7).

## 2.3. Statistical analysis

*Constructing the dataset:* We first built a dataset that consisted of growth probability as a function of tubulin concentration for all 12 configurations. To improve the statistical power of our approach, we added data for additional 12 configurations (four each in the growing, shortening, and transition phases). Once this process was complete, we had a data set in which growth/shortening probability is the response (i.e., the dependent variable), and the predictors (the independent variables) are the free tubulin concentration, the average number of GTP-bound subunits at the base layer of cracks, the average crack depth, and the number of GTP-bound subunits per PF.

*Initial considerations:* To obtain the quantitative relationship between the response and the predictors, we first needed to define our variables.

Let $\pi_{\text{grow}}$, $\pi_{\text{shorten}}$ and $\pi_{\text{transition}}$ be the probability that MT is in the state of growth, shortening and transition, respectively. Note that these probabilities are non-negative and satisfy $\pi_{\text{shorten}} + \pi_{\text{shorten}} + \pi_{\text{transition}} = 1$, since the MT is constrained to be in one of the three states. Let $X_{[\text{Tu}]}$, $X_{\text{ck-GTP}}$, $X_{\text{depth}}$ and $X_{\text{GTP}}$ be free tubulin concentration (in μM), the average number of GTP-bound subunits at the base layer of cracks (in subunits), the average crack depth (in subunits), and the number of GTP-bound subunits per PF, respectively. Given these variables, a naive way to model the probabilities would be a linear regression model

$$\pi_{\text{grow}} = \beta_0 + \beta_1 X_{[\text{Tu}]} + \beta_2 X_{\text{ck-GTP}} + \beta_3 X_{\text{depth}} + \beta_4 X_{\text{GTP}}$$

$$\pi_{\text{shorten}} = \beta_0' + \beta_1' X_{[\text{Tu}]} + \beta_2' X_{\text{ck-GTP}} + \beta_3' X_{\text{depth}} + \beta_4' X_{\text{GTP}}$$

$$\pi_{\text{transition}} = 1 - \pi_{\text{grow}} - \pi_{\text{shorten}}$$

(1)

where $\beta_0$ and $\beta_0{'}$ are intercepts, $\beta_i$s and $\beta_i{'}$s are coefficients, $i = 1, \ldots, 4$. Unfortunately, this model usually performs very poorly ([16]) as the probabilities do not follow a Gaussian distribution, and moreover, when free tubulin concentration ($X_{[\text{Tu}]}$) is very high or very low, the three probabilities can be negative or $> 1$, which is nonsensical. A commonly used alternative statistical model that overcomes these problems is the multinomial logistic model, which is a straightforward extension to the ordinary logistic model used for binary responses. This model is also called the baseline-category logit model and is described in detail in Chapter 7 of [16]. We chose this model for further analysis and describe it in detail below.

*A multinomial logistic model*: We proposed the following model, which treats $\pi_{\text{grow}}$, $\pi_{\text{shorten}}$, and $\pi_{\text{transition}}$ as the probabilities of a multinomial distribution with

$$\log\left(\frac{\pi_{\text{grow}}}{\pi_{\text{transition}}}\right) = \mu + \beta_1 X_{[\text{Tu}]} + \beta_2 X_{\text{ck-GTP}} + \beta_3 X_{\text{depth}} + \beta_4 X_{\text{GTP}}$$
$$\log\left(\frac{\pi_{\text{shorten}}}{\pi_{\text{transition}}}\right) = \mu' + \beta_1{'} X_{[\text{Tu}]} + \beta_2{'} X_{\text{ck-GTP}} + \beta_3{'} X_{\text{depth}} + \beta_4{'} X_{\text{GTP}}$$
$$\pi_{\text{transition}} = 1 - \pi_{\text{grow}} - \pi_{\text{shorten}}$$

(2).

The likelihood function of this model is still convex, so the estimation of the coefficients can be readily obtained by maximizing the likelihood.

We used our data set to fit the above multinomial logistic regression (Eqn. (2)) and estimated $\mu$ and $\mu'$, $\beta$s and $\beta'$s using the statistical software environment R [17] and its library "VGAM" [18].

To test the prediction power of our regression as shown in Supplementary Data Figures 3 and 4, we arbitrarily chose 16 new spontaneously occurring tip structures and used computational simulations to determine how their growth/shortening probability varied as a function of tubulin concentration; these trials were run by starting new simulations from these structures at a range of different tubulin concentrations. In parallel, we calculated for each of these structures the average number of GTP-bound subunits at the base layer of cracks, the average crack depth, and the number of GTP-bound subunits per PF, and then plugged the values of these predictors into Eqn. (2) to calculate the predicted growth/shortening probability. We then compared the predicted growth/shortening probabilities with those obtained from the computational simulations (results are shown in Supplementary Data Figures 3 and 4).

*Standardizing the coefficients of the model:* In order to compare the effects of these different variables in the multinomial logistic regression as described above ($X_{\text{ck-GTP}}$, $X_{\text{depth}}$, $X_{\text{GTP}}$, $X_{[\text{Tu}]}$), we standardized all these quantities so that they all have mean 0 and variance 1. The procedure of standardization is as follows. First, we calculate the mean for the aspect we want to standardize. For example, to standardize $X_{\text{ck-GTP}}$, we calculate its mean as $\mu_{\text{ck-GTP}}$. Then, we calculate the standard deviation for the aspect we want to standardize. Using $X_{\text{ck-GTP}}$ as an example, we calculate its standard deviation as $\sigma_{\text{ck-GTP}}$. Finally, for each data point, we use this information to calculate a new standardized data point. For example, suppose that the value of $X_{\text{ck-GTP}}$ is $x_{\text{ck-GTP}}$, we standardize $x_{\text{ck-GTP}}$ to $(x_{\text{ck-GTP}} - \mu_{\text{ck-GTP}})/\sigma_{\text{ck-GTP}}$. After these steps, we can calculate the new mean as $\mathrm{E}\left(\frac{X_{\text{ck-GTP}} - \mu_{\text{ck-GTP}}}{\sigma_{\text{ck-GTP}}}\right) = \mathrm{E}\left(\frac{X_{\text{ck-GTP}}}{\sigma_{\text{ck-GTP}}}\right) - \frac{\mu_{\text{ck-GTP}}}{\sigma_{\text{ck-GTP}}} = \frac{\mu_{\text{ck-GTP}}}{\sigma_{\text{ck-GTP}}} - \frac{\mu_{\text{ck-GTP}}}{\sigma_{\text{ck-GTP}}} = 0$ and the new variance as $\mathrm{Var}\left(\frac{X_{\text{ck-GTP}} - \mu_{\text{ck-GTP}}}{\sigma_{\text{ck-GTP}}}\right) = \frac{\mathrm{Var}(X_{\text{ck-GTP}} - \mu_{\text{ck-GTP}})}{\sigma_{\text{ck-GTP}}^2} = \frac{\sigma_{\text{ck-GTP}}^2}{\sigma_{\text{ck-GTP}}^2} = 1$. We perform these steps for each predictor so that all these four predictors have mean 0 and variance 1. Then, we can compare their effect on microtubule dynamics by comparing their coefficients in the regression model (Eqn. (3a) and Eqn. (4a)).

*Significance testing:* We tested the statistical significance of each term in Eqn. (2), that is, whether each of $\beta_i$ and $\beta_i'$ coefficients is significantly different from 0, using the likelihood ratio test, as follows. **STEP 1**: Calculate the deviance of full model, i.e. Eqn. (2) and denote this deviance as $D_1$. Deviance is a measure of goodness-of-fit in generalized linear models [16]. **STEP 2**: For each of the terms in Eqn. (2), generate a "reduced model" that includes all explanatory variables except the variable being tested, and calculate the deviance of that reduced model. For example, if the goal is to test the significance of the effect of the free tubulin concentration ($X_{[\mathrm{Tu}]}$), the reduced model includes all terms in Eqn. (2) except $\beta_1 X_{[\mathrm{Tu}]}$ and $\beta_1' X_{[\mathrm{Tu}]}$. We denote the deviance of a particular reduced model as $D_2$. **STEP 3**: Calculate the difference between the deviances $D_2 - D_1$. Statistically, $D_2 - D_1$ follows a $\chi^2$ distribution with two degrees of freedom. **STEP 4**: Calculate the p-value, which equals the right tail probability of the $\chi^2$ distribution. A small p-value means that the reduced model is significantly worse than the full model; in other words, the variable being tested is statistically significant and cannot be eliminated from the full model. As usual, we use 0.05 as the cutoff and claim an effect to be statistically significant if the p-value is less than 0.05.

## 2.4. Examining the correlation between polymerization behavior and the nucleotide state of the subunits at the bottom of the cracks in spontaneously occurring tip structures

To perform this analysis, we ran long simulations (corresponding to ~3 hours of simulated time) for free tubulin concentrations 8μM, 10μM, 11μM, 13μM. Then, we divided each simulation into 3 second time intervals, yielding ~3600 3s-intervals for each free tubulin concentration. For each of these intervals, we chose the structure at the middle point *t* , and determined the instantaneous growth velocity at time *t* by performing an ordinary linear regression of the MT lengths as a function of time in the time interval [*t*-1/2, *t*+1/2]. In such a way, we obtained ~3600 spontaneous structures and their associated instantaneous growth velocities. Consecutive structures are 3 seconds apart which decreases the correlation between consecutive structures. We plot instantaneous growth velocities of these 3600 structures vs. the fraction of GTP-bound subunits at the bottom of cracks (this fraction has a value of 1, 0.5, or zero corresponding to 2, 1, or zero GTP subunits) for each concentration in Fig 9.

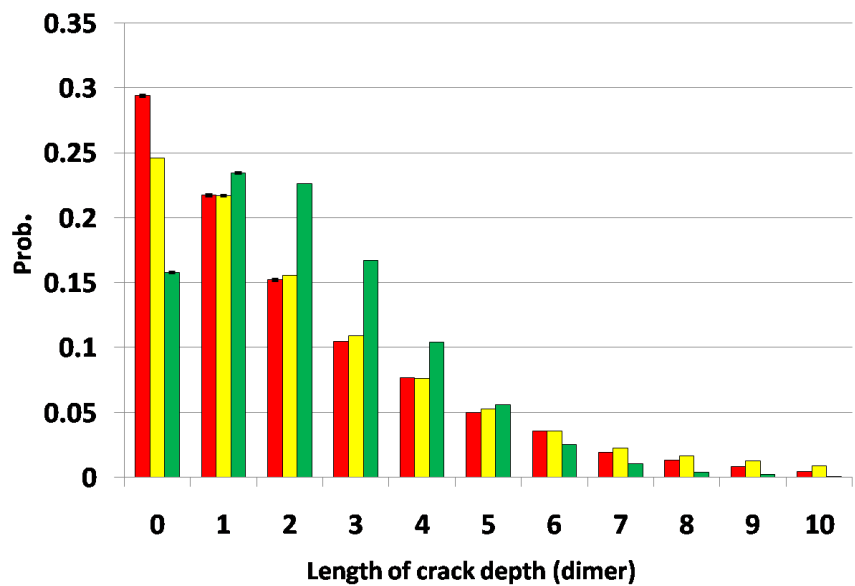

**A**

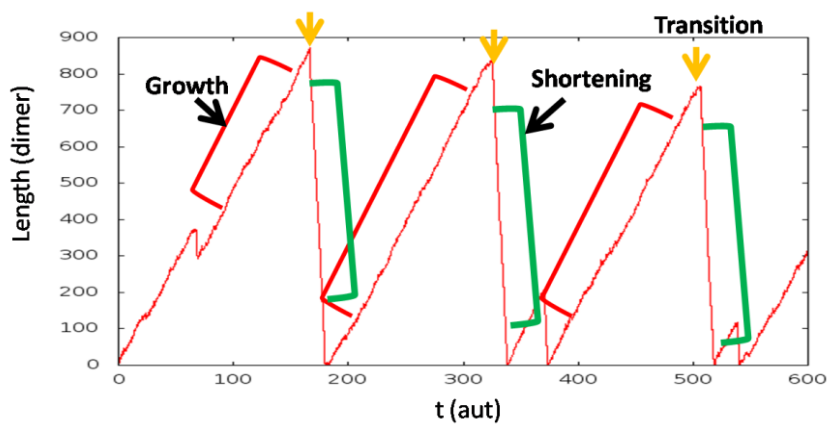

**B**

**Fig. 3.** Examination of the depth of spontaneously occurring cracks between protofilaments as they occurred in the reference simulation (**Fig. 2**). **A.** Distribution of the depth of cracks between protofilaments as observed during growth (red), shortening (green), and transition (yellow) phases. **B.** Time intervals used for the analysis. To create the plot in (**A**), we extracted all of the spontaneously occurring tip configurations that occurred in the initial simulation during the time intervals indicated in (**B**), then determined the average crack depth for each of these individual configurations. Then for each of the nine indicated time intervals, we created a histogram of the frequency at which each crack depth was observed. The histograms for the three intervals in each state (growth, shortening, or transition) were then averaged together to obtain the final histograms of the crack depth frequency. The error bars indicate the standard deviation; note that the value of this error is small because the number of structures is so large.

## 3. Results

### 3.1. Changes in crack depth have little effect on dynamic instability

We earlier concluded using a mean-field theory that the depth of the cracks between protofilaments correlated with dynamic instability. Therefore, as a first step, it was important to confirm this conclusion using detailed 13-protofilament computational model [11]. Fig. 3 shows the distribution of crack depths as observed in an initial simulation performed under our standard conditions (parameter set C from [11], 10 μM tubulin). These data show that the distribution of inter-protofilament crack depths in growing and depolymerizing MTs (red and green bars, respectively) is indeed different. In the shortening state, the histogram is shifted to the right (towards deeper cracks), indicating that as predicted, depolymerizing MTs have deeper cracks on average than do polymerizing MTs. These simulation data demonstrate that the crack depth does correlate with MT behaviour. However, we were puzzled that there was so little difference between MTs in the growing and transition states (compare the red and yellow bars).

To further investigate the influence of crack depth and test whether changes in crack depth simply correlate with or cause changes in DI, we used the 13PF simulation [11] to test the effect of reducing or extending the crack depth. More specifically, we first arbitrarily chose 12 spontaneously-occurring MT structures from the initial simulation: 4 MT structures each in the growth, shortening, and transition states (Fig. 2). Then, to test whether reducing the crack depth might promote growth, we "healed" all the cracks in the 4 shortening and 4 transitional MT structures, and used these altered structures to start a set of new simulations to determine how likely they were to grow, depolymerize, or pause. In parallel, we tested whether deepening the cracks could promote depolymerization by extending by three subunit lengths the cracks in the 4 growing and 4 transitional MT structures and using these altered structures to start new simulations. With these data in hand, we then compared the growth/shortening probabilities with and without the indicated alteration for each of the tested MT structures (Fig. 4)**.**

We obtained surprising results. We expected that healing of the interprotofilament cracks in depolymerizing or transitional MTs would promote MT growth but instead observed that it had little effect on the probability that either type of structure would grow at a given tubulin concentration (Fig. 4, top two rows). Similarly, we predicted that crack extension would promote MT depolymerization in growing or transitional MTs, but again observed little effect. More specifically, crack extension in growing MTs had little impact on the likelihood of continued MT growth (Fig. 4, third row), though crack extension in transitional structures did moderately increase the likelihood of depolymerization (Fig. 4, bottom row).

To explain these observations, we examined these results in more detail, and noticed that the ability of crack extension to promote depolymerization was greatest for the second of the four transitional structures that we examined. To investigate the possibility that this configuration had unusually large crack depths to start with, we plotted the maximum crack depth and the average crack depth of each of the 12 configurations (Fig. 5A, B). Examination of these data (Fig. 5A, B) shows that neither the average crack depth nor the maximum crack depth of the second transition configuration is larger than for other transition configurations. In fact, the average crack depth of this configuration is actually smaller than that of the other transitional configurations (Fig. 5B, purple bars). These data show that some feature of this structure other than crack depth is responsible for the increased sensitivity of this structure to additional extension of crack depth.

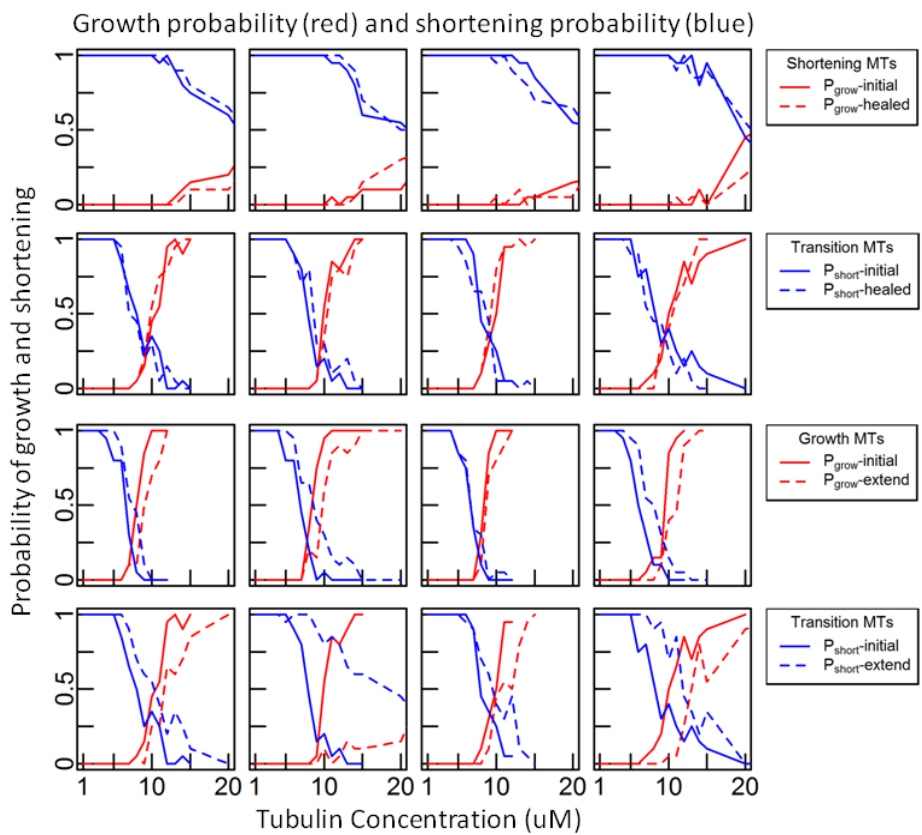

**Fig. 4.** Effect of changing crack depth on dynamic instability. The top two rows show the effect of healing all cracks between protofilaments for a single step, while the bottom two rows show the effect of extending all cracks by three subunit lengths for a single step. Each plot shows the probability of growth (red) or shortening (blue) as a function of tubulin concentration for a given reference structure (solid lines) or its altered version (dashed lines; structure is altered by healing or extended cracks as indicated). The four panels in each row correspond to the four reference structures from each of the three dynamic instability states (growth, shortening, or transition) as shown in Fig. 2; note that there are 16 panels instead of 12 because the transition structures were used in both crack healing and crack extension trials. These data show that in most cases, there is little effect of either crack healing or crack extension, at least within the parameter ranges tested.

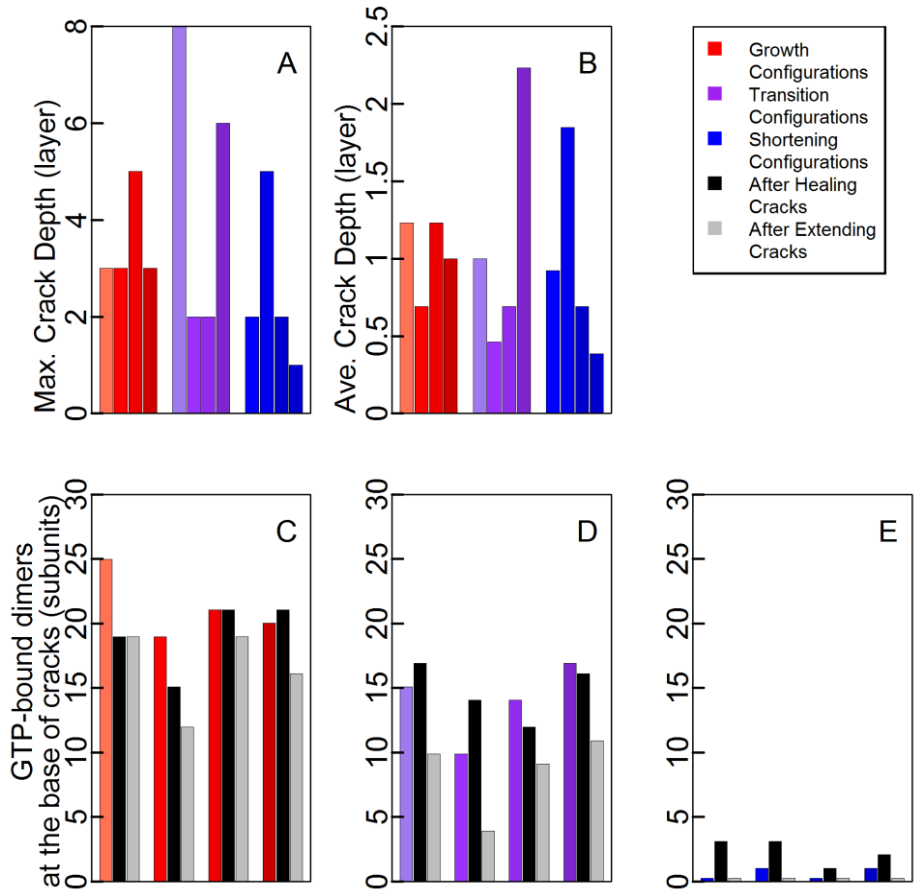

**Fig. 5.** **A-B.** Maximum (A) and average (B) crack depths of each of the initial configurations. The y-axis is the length of the deepest crack as measured in tubulin subunits. The four different shades of red, purple, and blue bars represent the four reference growth, transitional, and shortening configurations respectively. **C-E.** Number of subunits at the bases of the cracks that were GTP-bound in each of the initial configurations (colored as in (A)), after the cracks were healed (black bars), and after the cracks were extended (gray bars). The data for the growing (C), transitional (D) and shortening (E) configurations are presented in separate panels.

Taken together, these results indicate that while changes in crack depth do correlate with changes in microtubule behaviour, they do not directly cause changes to dynamic instability, at least within the parameter ranges tested. Given this conclusion, logic suggests that there does exist a crack depth sufficient to induce microtubule depolymerization. To test this prediction, we conducted additional simulations in which the cracks were extended by 9 subunits instead of 3, and found that tips with these hyper-extended cracks were indeed likely to depolymerize (Supplementary Data Figure 1). However, the low likelihood of cracks 9 or more subunits deep in spontaneously occurring structures (Fig. 3), together with the other data discussed above, suggested that some feature other than crack depth *per se* was involved in determining MT behaviour. In the next section we provide evidence that this "other feature" is the nucleotide state of the subunits (subunit pair) at the bottom of the cracks between protofilaments.

## 3.2. The nucleotide state of the subunits at the bases of the inter-protofilament cracks has a strong impact on dynamic instability

In Margolin et al. [11], we observed that catastrophe tended to occur when the cracks extended into GDP-rich regions. This observation suggested that the identity of the subunits at the bottom of cracks that matters most for dynamic instability, not the depth of the cracks *per se*. To test this idea, we plotted the percentage of subunits at the base of cracks that were GTP-bound for each of the 12 reference configurations (Fig. 2) before/after healing or extending the inter-protofilament cracks. These data are shown in Fig. 5C-E. The first thing we observed was that healing all cracks had little effect on the fraction of cracks that terminated at pairs of GTP-bound tubulin subunits, regardless of the growth/depolymerization state of the initial structure. Similarly, extending the cracks in growing structures by three subunits had relatively little effect on the fraction of cracks that terminated at GTP-bound tubulin subunits. However, a stronger effect was observed for transitional structures, and moreover, the reduction of GTP-terminated cracks was greatest for the second of the four transition structures examined (Fig. 5D). As noted above, this structure was most sensitive to crack extension (Fig. 4, bottom row). These observations are consistent with the hypothesis that the nucleotide state of the subunits at the base of the cracks is important for determining whether a microtubule is likely to keep growing, keep depolymerizing, or undergo transition from growth/shortening to shortening/growth. In addition, they provide a potential explanation for why the second transitional structure was the most sensitive to crack extension.

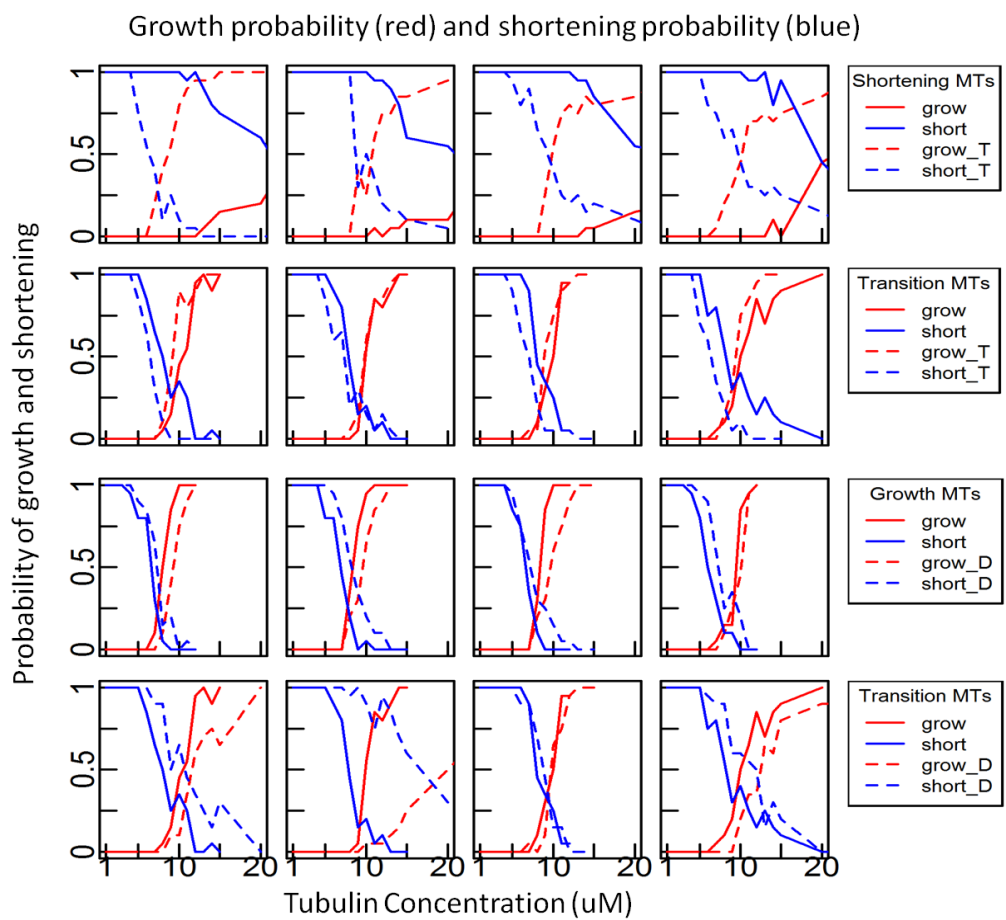

**Fig. 6.** Effect of changing the nucleotide state of the subunits at the bottom of the cracks. The top two rows show the effect of changing all crack-terminating subunits to GTP for a single step, while the bottom two rows show the effect of changing all crack-terminating subunits to GDP for a single step. Each plot shows the probability of growth (red) or shortening (blue) as a function of tubulin concentration for a given reference structure (solid lines) or its altered version (dashed lines; structure is altered by changing crack-terminating subunits to GTP or GDP as indicated). The four panels in each row correspond to the four reference structures from each of the three dynamic instability states (growth, shortening, or transition) as shown in Fig. 2; as in Figure 4, there are 16 panels instead of 12 because the transition structures were used in both sets of trials. These data show that changing the identity of the crack-terminating subunits has a dramatic effect on depolymerizing MTs, a small but detectable effect on growing MTs, and an intermediate effect on transitional MTs.

To investigate in more depth the idea that the nucleotide state (or more broadly, the conformation) of the subunits at the bottom of the cracks can influence the behavior of a MT tip, we ran a set of simulations similar to those in Fig. 4. However, instead of altering the crack depth, we altered for a single step the identity of the subunit pair at the base of each crack (changing any GDP subunits to GTP), and then examined the probability that this structure would give rise to a growing or shortening MT at various tubulin concentrations (Fig. 6). These data show that changing the identity of the crack-terminating subunits so that all are in the GTP conformation had a dramatic effect on depolymerizing structures - it shifted all towards growth promotion (Fig. 6, top row). The effect of performing similar operations on transitional structures was less dramatic but still detectable: after this single-step change, the transitional structures were more likely to grow. While this change seems small, it is important to point out that the altered transitional structures behaved in a way that was almost indistinguishable from growth structures (Fig. 6, compare the second and third rows). Performing the opposite operation, i.e., changing crack-terminating subunits from GTP to GDP tubulin, had a detectable but not dramatic effect on growing MTs (Fig. 6, third row), making them more likely to depolymerize. This operation had a stronger effect on transitional MTs, although the strength of the effect varied dramatically with the particular configuration tested (Fig. 6, last row).

In considering these results, it is important to remember that when we altered the structures by changing the crack depth (Fig. 4) or the identity of the crack-terminating subunits (Fig. 6), we changed the structures only in the starting configuration; after this initial step, the rest of the simulation ran according to standard rules. Moreover, for Fig. 6, the only GDP subunits that were changed to GTP (or *vice versa*) were those that were part of the subunit pair at the very base of a crack - the rest of the subunits in the simulated MT remained unchanged. These simulation results suggest that the identity of the subunits at the base layer of cracks is indeed significant for the dynamic instability of MTs.

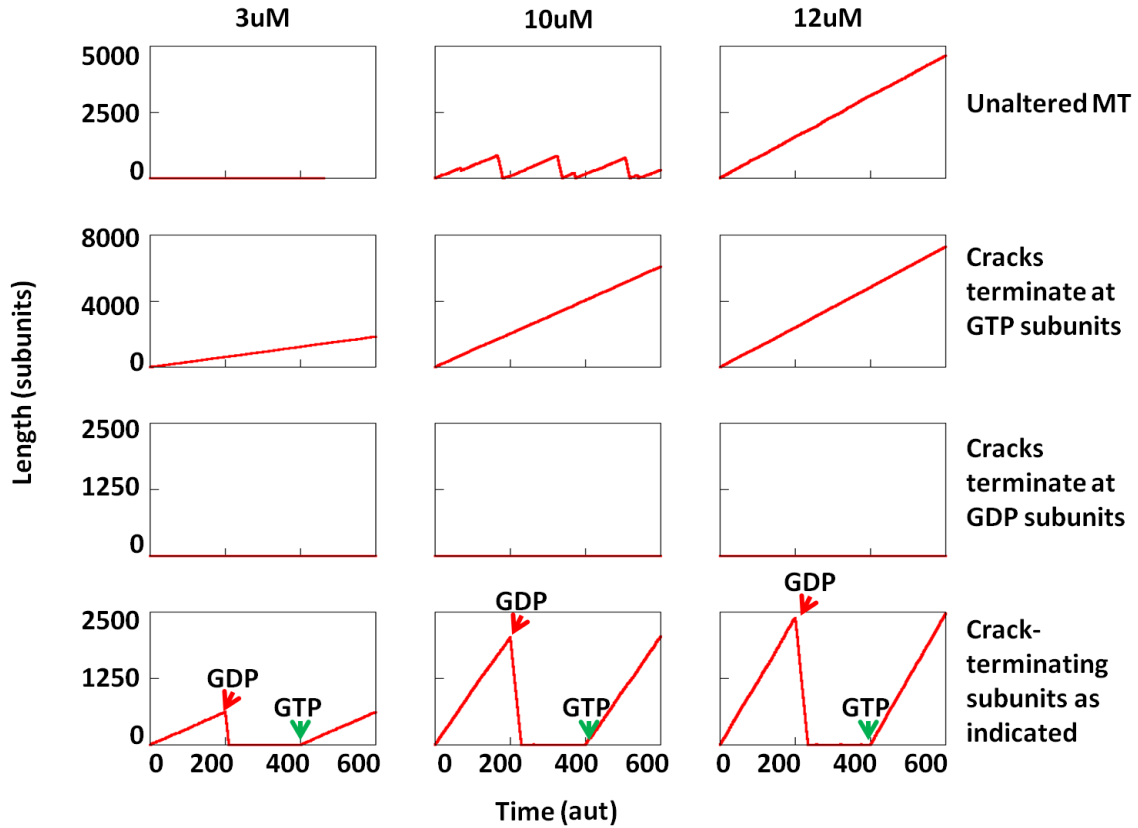

**Fig. 7.** Effect of changing the nucleotide state of crack-terminating subunits constantly throughout the course of a simulation. The top row shows the behavior of MTs run under the standard simulation conditions at tubulin concentrations as indicated. The second and third rows show the behavior of MTs where all crack-terminating subunits are instantaneously changed to GTP or GDP (respectively) at each simulation step. The fourth row shows a set of simulations where the crack-terminating subunits are set to GTP as in the second row, then changed to GDP and back to GTP as indicated. Together with Fig. 6, these data show that the identity of the crack-terminating subunits can have a profound effect on dynamic instability.

To test this hypothesis further, we examined what would happen if instead of changing the identity of the crack-terminating subunits for a single step, we altered them for the duration of the simulation. We also tested the dependence of the observed behavior on free tubulin concentration (Fig. 7). These results show that the identity of the crack-terminating subunits has a profound influence on the behavior of the simulated MTs: changing the crack-terminating subunits to GTP is sufficient to allow

MTs to grow at 3 μM tubulin, while changing the crack-terminating subunits to GDP is sufficient to prevent MTs from growing at 12 μM tubulin.

These observations give strong support for the idea that the nucleotide state of the crack-terminating subunits plays a controlling role in dynamic instability. However, we realized that there was a caveat: by constantly changing the identity of the crack-terminating subunits and otherwise letting the simulation run normally, we were indirectly altering the number of GTP tubulin subunits in each tip. One could argue that the effects on dynamic instability observed in Fig. 7 were simply a result of changing the size of the GTP cap. To investigate this possibility, we examined the number of GTP tubulins in the growing tips under the various conditions used in Fig. 7. These data (Table 1 and Fig. 8) show that changing all crack-terminating GDP subunits to GTP does cause a perceptible increase in the size of the GTP cap, as measured by the number of GTP subunits in the tip. However, closer inspection shows something quite interesting: the manipulated tips grow persistently at 3 μM tubulin, even though they have smaller GTP caps than do the normal tips at 10 μM tubulin, which undergo frequent catastrophes (Figs. 7-8). Therefore, we conclude that the identity of the crack-terminating subunits does have a significant impact on dynamic instability, and that this influence can be separated from the influence of the overall size of the cap.

### 3.3. Quantitative relationship between growth/shortening probability and free tubulin concentration, number of GTP-bound subunits and crack depth

In this section, we use statistical approaches to determine the quantitative relationship between growth/shortening probability and the following characteristics of the MT tip structures that occur spontaneously in our simulations:

a) average number of GTP-bound subunits at the bottom of each inter-protofilament crack (this is a number between 0 and 2), $X_{\text{ck-GTP}}$

b) average crack depth (this is a number between 0 and an unlimited value,

although Fig. 3 shows that it is unlikely to be greater than 10), $X_{\text{depth}}$

c) number of GTP-bound subunits per PF (this is a number between 0 and an unlimited value), $X_{\text{GTP}}$

d) free tubulin concentration (measured in micromolar), $X_{[\text{Tu}]}$

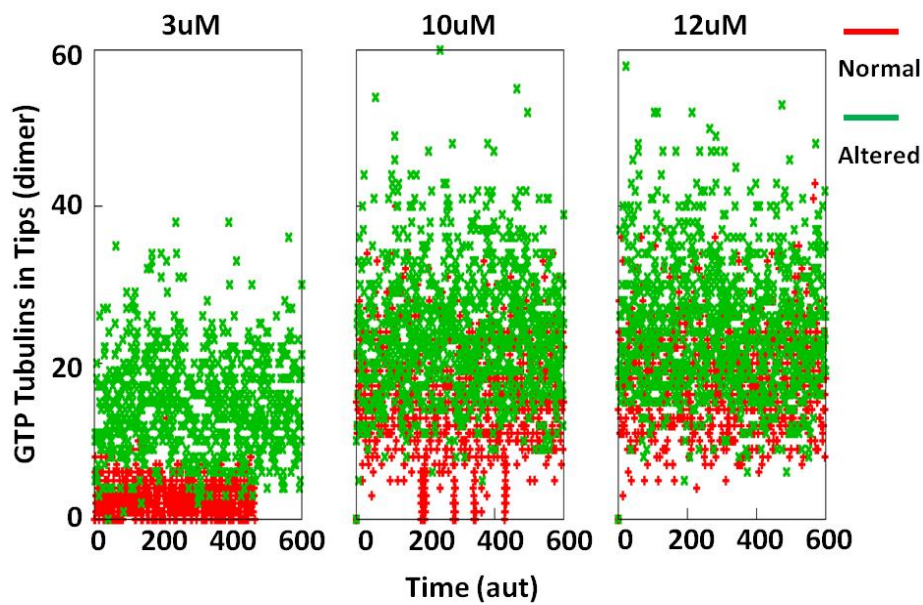

**Fig. 8.** Effect of manipulating the identity of the crack-terminating subunits on the size of the GTP cap. These panels show the number of GTP tubulins in the MT tip through the course of the simulations shown in the second row of Fig. 7. The red points correspond to the unaltered simulations, while the green points correspond to the simulations where all crack-terminating subunits are changed to GTP tubulin. These data show that consistently changing the identity of crack-terminating subunits to GTP does increase the size of the GTP cap. However, it is important to note that the cap size of the altered MT at 3 μM tubulin is still smaller than the cap of the normal MT at 10 μM. Since the altered MTs grow persistently at 3 μM tubulin (Fig. 7), and normal MTs are dynamic at 10 μM tubulin, these data show that that the strong effect of altering the crack-terminating subunits does not result simply from increasing the size of the GTP cap.

Table 1. Number of GTP tubulins in the growing tips

|  | 3μM | 10μM | 12μM |
|---|---|---|---|
| Normal simulation | no growth observed | 16 (+/-0.85) | 18.34 (+/-1.15) |
| Manipulated simulation (All crack-terminating subunits changed to GTP) | 14 (+/-0.72) | 24.16 (+/-1.4) | 24.82 (+/-1.39) |

In performing this work, we hoped to achieve three goals. First, we wanted to see if we could find statistical support for the idea that the nucleotide state of the crack-terminating subunits is important for MT behavior. Second, we wanted to see if the statistical approach would allow us to quantitatively measure the importance of the identity of the crack-terminating subunits relative to other characteristics such as crack-depth and number of GTP subunits. Finally, we hoped to arrive at a predictive model that would be able to take an arbitrary MT tip structure and predict its propensity for growth or shortening.

**Multinomial logistical regression formula**

A standard approach for building a model relating characteristics such as these to an output response of one of three states (e.g., growth, shortening, or in transition) is multinomial logistic regression [16]. Briefly, as discussed more in Materials and Methods, we fit Eqn. (2) to our dataset of tip structures and related growth propensities at various tubulin concentrations, and extracted the coefficients for each term $X$ using the statistical software environment R. Our initial expectation had been that we would arrive at one regression formula regardless of the behavior of the MT at the time the tip structure was extracted, but it turned out that this was not realistic because the relationship between growth propensity and tubulin concentration is very different for tip structures extracted from growing and shortening MTs (Supplementary Data Figure 2). Therefore, we derived different formulas for the two major dynamic states of growth and depolymerization. Using multinomial logistic regression, we obtained the following regression formula for tip configurations that occurred during growth:

$$\log\left(\frac{\pi_{\text{grow}}}{\pi_{\text{transition}}}\right) = -4.6 + 4.82X_{[\text{Tu}]} + 0.82X_{\text{ck-GTP}} + 0.12X_{\text{depth}} - 0.85X_{\text{GTP}}$$

$$\log\left(\frac{\pi_{\text{shorten}}}{\pi_{\text{transition}}}\right) = 1.27 - 3.10X_{[\text{Tu}]} + 0.036X_{\text{ck-GTP}} - 0.24X_{\text{depth}} + 0.022X_{\text{GTP}}.$$

$$\pi_{\text{transition}} = 1 - \pi_{\text{grow}} - \pi_{\text{shorten}}$$

(3a)

$$\log\left(\frac{\pi_{\text{grow}}}{\pi_{\text{transition}}}\right) = -15.22 + 2.02X_{[\text{Tu}]} + 3.71X_{\text{ck-GTP}} + 8.64X_{\text{depth}} - 1.56X_{\text{GTP}}$$

$$\log\left(\frac{\pi_{\text{shorten}}}{\pi_{\text{transition}}}\right) = 9.81 - 1.19X_{[\text{Tu}]} - 0.09X_{\text{ck-GTP}} + 2.37X_{\text{depth}} - 0.39X_{\text{GTP}}.$$

$$\pi_{\text{transition}} = 1 - \pi_{\text{grow}} - \pi_{\text{shorten}}$$

(3b)

Eqn. (3a) is obtained using standardized data. Eqn. (3b) is obtained using original non-standardized data (see Materials and Methods for discussion of how we performed the standardization). We will use coefficients in Eqn. (3a) to compare the effects of different predictors.

The regression formula that we obtained for tip configurations extracted from shortening MTs is:

$$\log\left(\frac{\pi_{\text{grow}}}{\pi_{\text{transition}}}\right) = -1.29 + 2.16X_{[\text{Tu}]} + 0.4X_{\text{ck-GTP}} + 0.14X_{\text{depth}} - 0.41X_{\text{GTP}}$$

$$\log\left(\frac{\pi_{\text{shorten}}}{\pi_{\text{transition}}}\right) = 1.98 - 1.07X_{[\text{Tu}]} + 0.21X_{\text{ck-GTP}} + 0.089X_{\text{depth}} - 0.19X_{\text{GTP}}.$$

$$\pi_{\text{transition}} = 1 - \pi_{\text{grow}} - \pi_{\text{shorten}}$$

(4a)

$$\log\left(\frac{\pi_{\text{grow}}}{\pi_{\text{transition}}}\right) = 7.17 + 0.23X_{[\text{Tu}]} - 182.96X_{\text{ck-GTP}} - 27.69X_{\text{depth}} + 97.61X_{\text{GTP}}$$

$$\log\left(\frac{\pi_{\text{shorten}}}{\pi_{\text{transition}}}\right) = 10.41 - 0.05X_{[\text{Tu}]} - 105.5X_{\text{ck-GTP}} - 15.93X_{\text{depth}} + 50.59X_{\text{GTP}}.$$

$$\pi_{\text{transition}} = 1 - \pi_{\text{grow}} - \pi_{\text{shorten}}$$

(4b)

Eqn. (4a) is obtained using standardized data. Eqn. (4b) is obtained using original non-standardized data. We will use coefficients in Eqn. (4a) to compare the effects of different predictors.

**Statistical Inference**

For configurations that occurred during growth (Eqn. (3a)), the values of the coefficients of $X_{\text{ck-GTP}}$ and of $X_{\text{GTP}}$ are ~six times larger than the coefficient of $X_{\text{depth}}$ in $\log\left(\frac{\pi_{\text{grow}}}{\pi_{\text{transition}}}\right)$ part in Eqn. (3a) and they are ~six times smaller than the coefficient of $X_{\text{depth}}$ in $\log\left(\frac{\pi_{\text{shorten}}}{\pi_{\text{transition}}}\right)$ part in Eqn. (3a). This observation initially suggests that the influence on dynamic instability of the average number of GTP-bound subunits at the base layer of the cracks is similar to the influence of the average crack depth. However, it is important to determine whether the contribution of each of these terms (their effect) is statistically significant before interpreting these values. As shown in Supplementary Data Table 1, each of these terms does contribute to the fit in a statistically significant way (p-value < 0.05), indicating that all these three variables $X_{\text{ck-GTP}}$, of $X_{\text{depth}}$ and of $X_{\text{GTP}}$ effects are significant to MT behavior. As would be expected, free tubulin concentration is also significant for our response, (growth/shortening probability). Since the tubulin concentration is usually relatively high ($> 8\,\mu\text{M}$) in these simulations, the large size of this term relative to the others indicates that the dominant factor for the behavior of growing MT tips is the concentration of GTP tubulin. Thus, while all variables do contribute to MT behavior in a statistically significant way, it is difficult to discern quantitatively the relative contributions of the other variables in the presence of changes to soluble tubulin concentration (we address this issue further below).

With regard to the shortening state (Eqn. (4a)), Table 1 in Supplementary data shows that each of the three terms ($X_{\text{ck-GTP}}$, $X_{\text{depth}}$ and $X_{\text{GTP}}$) contributes to MT behavior in a statistically significant way. Examination of the coefficients of Eqn. (4a) shows that the magnitude of the coefficients of $X_{\text{ck-GTP}}$ and $X_{\text{GTP}}$ are ~2 times the magnitude of the coefficient of $X_{\text{depth}}$ in both $\log\left(\frac{\pi_{\text{grow}}}{\pi_{\text{transition}}}\right)$ and $\log\left(\frac{\pi_{\text{shorten}}}{\pi_{\text{transition}}}\right)$ parts in Eqn. (4a). These observations mean that both the average number of GTP-bound subunits at the base layer of the cracks and the number of GTP-bound subunits per PF influence MT dynamics more than does the average crack depth. Of course, the concentration of free tubulin also contributes, but in a less lop-sided way than with growing MTs, allowing the relative impact of the other terms to be detected.

As noted above, the strong impact of the free tubulin concentration on the behavior of growing MTs made it difficult to detect the relative importance of the different terms (crack depth vs. identity of the crack-terminating subunits, etc.). Therefore, we performed multinomial logistic regression for these variables at fixed free tubulin concentration in order to provide the magnitude of influence of these three factors. The detailed multinomial regression equations are in supplementary data (Eqn. (A1) - Eqn. (A4)). From the coefficients of these equations, we can see that average number of GTP-bound subunits at the bottom of the cracks and the average crack depth have stronger effects on DI than does the number of GTP-bound subunits per PF. The p-values of $X_{\text{ck-GTP}}$ and $X_{\text{depth}}$ are smaller than 0.05 at all free tubulin concentrations tested (8μM -11μM, Table 2 in Supplementary Data), which indicates that they are significant for DI. It is interesting to note that while the p-value of $X_{\text{GTP}}$ is smaller than 0.05 at some free tubulin concentrations (10μM, 11μM), it is larger than 0.05 at others (8μM, 9μM). This value implies that average number of GTP-bound subunits per PF is not always significant for DI. The observations lead us to conclude that the number of GTP-bound subunits impacts DI indirectly, through its influence on the likelihood that cracks terminate at pairs of GTP subunits. This conclusion is consistent with the results of our simulation study.

### 3.4. Examining the correlation between polymerization behavior and the

**nucleotide state of the subunits at the bottom of the cracks in spontaneously occurring tip structures**

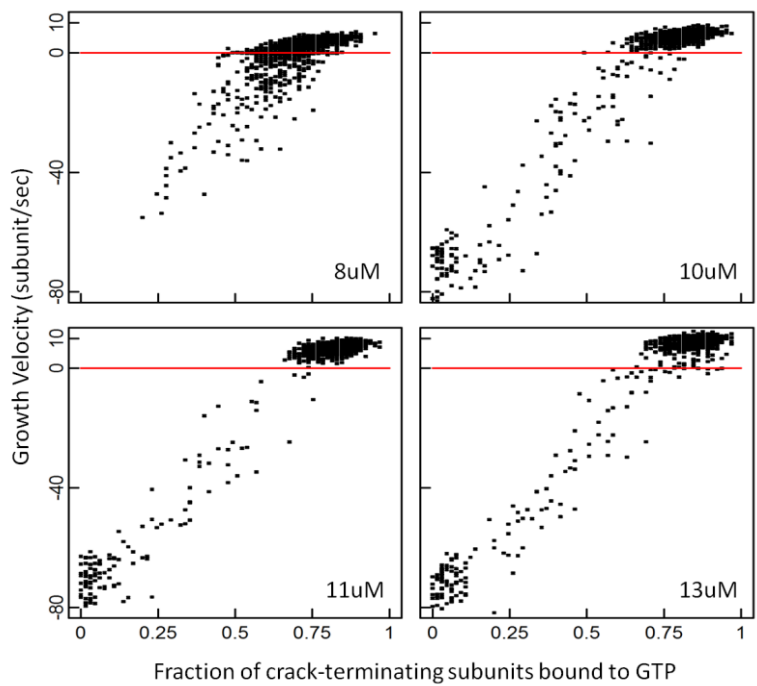

**Fig. 9.** The relationship between growth velocity and the fraction of GTP-bound subunits at the bottom of cracks. The free tubulin concentrations are 8μM (top left), 10μM (top right), 11μM (bottom left) and 13μM (bottom right). The x-axis is the fraction of GTP-bound subunits at the bottom of cracks. The y-axis is the growth velocity in subunit/sec. Many points are very close so each black dot on the plot probably represents many points. The relationship between growth velocity and the fraction of crack-terminating subunits bound to GTP is almost linear when free tubulin concentration is ≥ 10μM. When free tubulin concentration is 8μM, microtubule cannot grow long and behave noisily. Microtubule under 8uM free tubulin concentration can depolymerize completely with less than 3 seconds which is the time interval we choose. Most of 3 second time interval covers the growth periods, which leads that we do not see many dots close to the left bottom corner in the plot of 8μM.

The data discussed above provided strong support for the idea that the nucleotide state of the crack-terminating subunits plays a key role in determining the growth propensity of a given tip structure. However, our analysis was based on a relatively small number of tip structures. To examine the generality of the relationship, we ran long simulations (3 hours of simulated time) at each of four tubulin concentrations. We identified tip structures at one-second intervals, and measured the instantaneous growth velocity of the MT at each of these intervals (Fig. 9). These data show that having a high fraction of GTP subunits at the base of the cracks (fraction between 0.8 and 1) correlates strongly (nearly completely) with microtubule growth. The correlation coefficients between growth velocity and fraction of crack-terminating GTP-bound subunits at the bottom of the cracks are 0.72, 0.95, 0.97, 0.97, at 8μM, 10μM, 11μM, 13μM, respectively.

## 4. Conclusions

In this paper, we have used a detailed computational model of MT dynamics to elucidate the relationship between the structure of the MT tip and dynamic instability behavior. One specific goal was to investigate the role of cracks between protofilaments, since our previous work with a mean field model of dynamic instability had implied that the depth of these cracks plays a key role in determining whether a MT tip is likely to grow or depolymerize [12]. Consistent with this work, we found that altering the depth of the inter-protofilament cracks in otherwise identical MT tips does influence MT behavior. However, further investigation demonstrated that it is not the depth of the cracks *per se* that matters in determining whether a MT is likely to grow, shrink, or undergo transition, and led us to hypothesize that what matters most is whether the cracks terminate in the regions of GTP or GDP-rich tubulin.

To test this hypothesis in the context of our simulation, we examined the effect of changing the nucleotide state of the crack-terminating subunits, and showed that this perturbation had a strong effect on MT dynamics, one that could be separated from its effect on the GTP cap. To further study this idea, we examined the relationship between structure of the tip and its growth propensity through multinomial logistic regression, and found that the nucleotide state of the crack-terminating subunits is indeed a significant predictor of MT behavior. However, the number of GTP subunits in the tip, and the depth of the cracks were also significant predictors. How can this observation be reconciled with the idea that the identity of the crack-terminating subunits plays a dominant role? We suggest that the size of the GTP cap and the depth of the cracks are also significant predictors because when the cap is deep and the cracks are shallow, fluctuations in the depth of the cracks are less likely to lead to crack extension into GDP-rich regions.

These observations suggest that a growth promoting "GTP cap" is one where the cracks terminate in pairs of GTP-bound subunits. This definition is relevant only to a particular moment in time, since the MT tip fluctuates rapidly. However, it follows that MTs will tend to undergo period of extended growth under conditions where the cracks are likely to terminate in pairs of GTP-bound subunits, and unlikely to terminate in GDP subunits. Therefore, we propose that a functionally significant (growth promoting) GTP cap is, on average, one where it is unlikely that cracks extend into GDP-rich regions. Thus, the depth (size) of the GTP cap is also important because when the cap is deep, fluctuations in the depth of the cracks are less likely to lead to crack extension into GDP-rich regions. This idea means that the size and shape of a growth promoting GTP cap will vary with the biochemical characteristics of the tubulin subunits, and it will also vary with the concentration of a given type of tubulin. In addition to helping to clarify the mechanism of dynamic instability, this idea has relevance for the mechanism of MT stabilizers: proteins that introduce lateral cross-links between protofilaments would produce islands of GDP tubulin that mimic GTP-rich regions in having strong lateral bonds, reducing crack propagation, suppressing catastrophe and promoting rescue.

**Acknowledgements**

The research was supported by NSF grants MCB-0951264 and MCB-1244593 to HVG and MSA.